\documentclass{Interspeech}
\usepackage[table,xcdraw]{xcolor}
\usepackage{booktabs}
\usepackage{multirow}

\interspeechcameraready

\title{
    VoxAging: Continuously Tracking Speaker Aging with a Large-Scale Longitudinal Dataset in English and Mandarin
}

\author[affiliation={1}]{Zhiqi}{Ai}
\author[affiliation={1}]{Meixuan}{Bao}
\author[affiliation={1}]{Zhiyong}{Chen}
\author[affiliation={1}]{Zhi}{Yang}
\author[affiliation={2}]{Xinnuo}{Li}
\author[affiliation={3,*}]{Shugong}{Xu}

\affiliation{}{Shanghai University}{China}
\affiliation{}{New York University}{USA}
\affiliation{}{Xi’an Jiaotong-Liverpool University}{China}
\email{aizhiqi-work@shu.edu.cn, shugong.xu@xjtlu.edu.cn}

\usepackage{comment}

\begin{document}

\maketitle
\renewcommand{\thefootnote}{\fnsymbol{footnote}}
\footnotetext[1]{Corresponding author}
\renewcommand{\thefootnote}{\arabic{footnote}}

\begin{abstract}

    The performance of speaker verification systems is adversely affected by speaker aging. However, due to challenges in data collection, particularly the lack of sustained and large-scale longitudinal data for individuals, research on speaker aging remains difficult. In this paper, we present VoxAging, a large-scale longitudinal  
    dataset collected from 293 speakers (226 English speakers and 67 Mandarin speakers) over several years, with the longest time span reaching 17 years (approximately 900 weeks). For each speaker, the data were recorded at weekly intervals. We studied the phenomenon of speaker aging and its effects on advanced speaker verification systems, analyzed individual speaker aging processes, and explored the impact of factors such as age group and gender on speaker aging research.

\end{abstract}

\section{Introduction}

Speaker recognition (SR) and face recognition (FR) are widely used biometric technologies for identity authentication \cite{zhao2003face, kabir2021survey}. However, both face challenges related to aging \cite{10599875, kelly2012speaker, baruni2021age}. As people age, physiological changes in the face and vocal tract lead to gradual alterations in their features, negatively affecting the accuracy of SR and FR systems. In SR systems, the impact of aging is particularly significant \cite{10599875, kelly2012speaker, kelly2015evaluation, kelly2016score, qin22_interspeech, singh23d_interspeech}. Aging affects the vocal cords and vocal tract, causing voiceprint features to deteriorate, which reduces the reliability of SR systems \cite{REUBOLD2010638, lei2009role}. Consequently, SR systems require more frequent updates to ID templates to maintain performance, as voiceprint features are highly sensitive to aging-related changes.

Early research on speaker aging was limited by scarce data and the capabilities of SR models. These studies primarily relied on traditional speech datasets with short time spans \cite{lei2009role, ramig1983effects, reynolds2000speaker}. For instance, \cite{ramig1983effects} used SEARP pitch analysis and observed that healthy individuals exhibited less tremor during vowel production, whereas elderly individuals displayed more pronounced tremors. Similarly, \cite{lei2009role} and \cite{reynolds2000speaker} employed models such as GMM-UBM and found that speaker aging negatively impacted SR system performance, suggesting that incorporating age-related factors could enhance accuracy.

\begin{figure}[t]
  \centering
  \includegraphics[width=\linewidth]{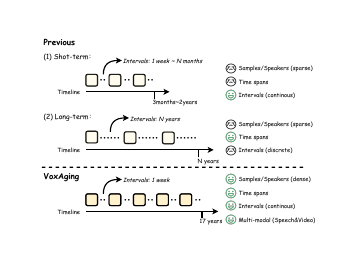}
  \caption{Previous short-term datasets have continuous intervals but limited time spans, while long-term datasets have long time spans with discrete intervals, both with sparse sampling. The VoxAging offers dense sampling, continuous weekly intervals, long time spans, and multi-modal data.}
  \label{fig:datacom}
\end{figure}

Recent studies have increasingly focused on the impact of speaker aging using cross-age speaker datasets. Research on the TCDSA dataset \cite{kelly2012speaker, kelly2015evaluation, kelly2016score} shows that verification scores decline as the time span increases, with short-term aging effects being relatively minor. Over time, genuine speaker scores decrease significantly, while impostor scores remain stable \cite{kelly2015evaluation}. A fixed decision threshold can exacerbate classification error rates even with just a few years' age difference \cite{kelly2016score}. Recent work on advanced SR models, such as ResNet34 and ECAPA-TDNN \cite{qin22_interspeech, singh23d_interspeech}, confirms that aging-related changes degrade system performance. These effects are more pronounced in female English speakers but have a greater impact on male Finnish speakers \cite{singh23d_interspeech}.

A major challenge in speaker aging research is the scarcity of long-term data. Most existing datasets cover relatively short periods (typically around 3 months to 2 years) with a limited number of speakers, leading to data jitter and outliers. For instance, the CSLT-Chronos dataset \cite{wang2016improving} includes 60 speakers and 84,000 samples collected over two years. Long-term datasets, such as TCDSA, contain recordings from 17 speakers over a span of 28 to 58 years but with fewer than 10 samples per speaker \cite{kelly2012speaker}. The LCFSH Finnish dataset \cite{singh23d_interspeech} only covers two discrete intervals (20 and 40 years), while the VoxCeleb dataset \cite{qin22_interspeech, singh23d_interspeech}, annotated with age-related face models for aging analysis, still lacks sufficiently dense audio evaluation data for each speaker.

\begin{table*}[!ht]
\centering
\caption{Comparison of existing speaker aging datasets. '-' indicates unavailable information. "Discrete" means datasets with long session intervals, where each ID has only a few samples. "Continuous" means datasets with short session intervals and continuous collection. "gradient*" indicates that the session intervals gradually increase over time.}
\resizebox{\linewidth}{!}{
\begin{tabular}{@{}lcccclll@{}}
\toprule
Dataset         & \# of Spks & \# of Segments & \# of Hours & \# Max Span (years) & \# Session Intervals & Language          & Modality      \\ \midrule
\multicolumn{8}{l}{{\color[HTML]{C0C0C0} \textit{Discrete}}}                                                                       \\ \midrule
TCDSA \cite{kelly2012speaker}           & 17         & 231            & 30          & 58                  & 1$\sim$23 years      & English           & Speech        \\
LCFSH \cite{singh23d_interspeech}           & 109        & 15,474         & -           & 40                  & 20 years             & Finnish           & Speech        \\
VoxCeleb-AE \cite{singh23d_interspeech}     & 670        & 79,063         & \textless 352 & 10                  & -                    & English           & Speech        \\ 
VoxCeleb-CA \cite{qin22_interspeech}     & 971        & 92,635              & \textless 352 & 20                  & -                    & English           & Speech        \\ \midrule
\multicolumn{8}{l}{{\color[HTML]{C0C0C0} \textit{Continous}}}                                                                                       \\ \midrule
MARP \cite{lawson09_interspeech}            & 60         & -              & -           & 3                   & 2 months             & English           & Speech        \\
CSLT-Chronos \cite{wang2016improving}    & 60         & 84,000         & 70          & 2                   & gradient*            & Mandarin          & Speech        \\
SMIIP-TV \cite{qin2024investigating}        & 373        & 325,049        & 305         & 0.25                & 4 days               & Mandarin          & Speech        \\ \midrule
\rowcolor[HTML]{EFEFEF} 
VoxAging (Ours) & 293        & 2,629,100      & 7,522       & 17                  & 1 week               & English, Mandarin & Speech, Video \\ \bottomrule
\end{tabular}
}
\label{tab:cap}
\end{table*}

To address the challenges of speaker aging in SR systems, we present VoxAging, a large-scale longitudinal 
dataset. It includes recordings from 293 speakers (226 English and 67 Mandarin) over a span of 17 years, totaling 7,522 hours, with weekly samples. Our research investigates how aging affects voice features and the performance of advanced SR models, as well as the impact of age group and gender on speaker aging.


\section{VoxAging Dataset}
\subsection{Previous speaker aging datasets}

As shown in Table \ref{tab:cap}, existing speaker aging datasets can be classified into two types: discrete and continuous, based on session intervals. Discrete datasets \cite{kelly2012speaker, singh23d_interspeech, qin22_interspeech} have long session intervals and limited samples per speaker, spanning several years to two decades. For instance, TCDSA \cite{kelly2012speaker} includes recordings from 17 speakers over a span of 28 to 58 years, but with fewer than 10 samples per speaker. LCFSH \cite{singh23d_interspeech}, a Finnish dataset, has only two time spans: 20 and 40 years. VoxCeleb-AE \cite{singh23d_interspeech} and VoxCeleb-CA \cite{qin22_interspeech}, derived from the VoxCeleb \cite{nagrani17_interspeech} dataset (originally designed for general speaker recognition), feature imprecise age labels and limited samples per speaker (an average of 123 utterances).

In contrast, continuous datasets \cite{lawson09_interspeech, wang2016improving, qin2024investigating} feature shorter session intervals and higher collection frequencies, ranging from a few months to days. MARP \cite{lawson09_interspeech} covers 60 speakers with a 2-month interval, CSLT-Chronos \cite{wang2016improving} includes 60 speakers over 2 years, with 14 sessions collected at gradient intervals, and SMIIP-TV \cite{qin2024investigating}, a recently collected dataset, tracks data from 373 individuals continuously over 3 months at a high cost.

\subsection{Data description}

The VoxAging dataset is a large-scale, longitudinal 
collection compiled from 293 speakers, including 226 English speakers (112 female, 114 male) and 67 Mandarin speakers (23 female, 44 male). The dataset spans up to 17 years (approximately 900 weeks) with weekly recordings, offering dense sampling over an extended period. It contains 2,629,100 segments, amounting to 7,522 hours of audio-visual data. The data was sourced from YouTube\footnote{\url{https://www.youtube.com}} and Bilibili\footnote{\url{https://www.bilibili.com}}, with channels manually filtered to ensure high-quality videos and appropriate time spans. As shown in Table \ref{tab:cap}, the unique advantage of the VoxAging dataset lies in its continuous weekly intervals over such an extended period, setting it apart from previous speaker aging datasets, which have limited time spans or discrete intervals.

\begin{figure}[t]
  \centering
  \includegraphics[width=\linewidth]{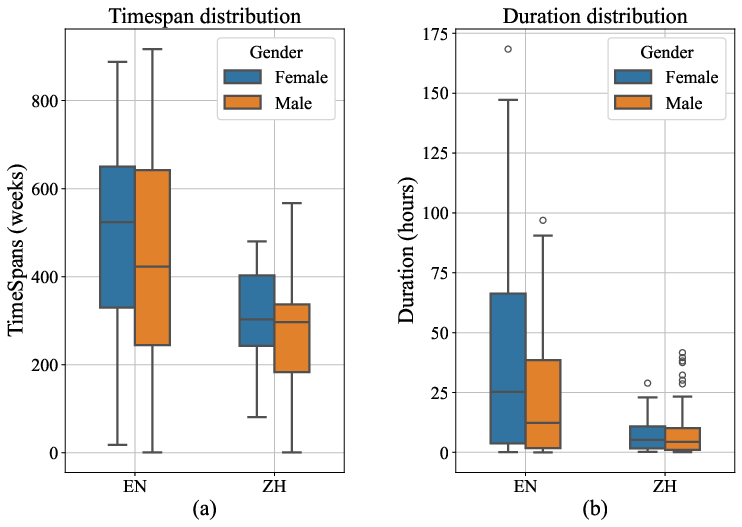}
  \caption{VoxAging dataset distribution: (a) timespan distribution, (b) duration distribution. In VoxAging, there are 293 speakers: 226 English speakers (112 female and 114 male) and 67 Mandarin speakers (23 female and 44 male).}
  \label{fig:eps1}
\end{figure}

Figure \ref{fig:eps1} illustrates the static distribution of the VoxAging dataset. The time span and data size for English speakers are larger, primarily because Mandarin data collection is more challenging, with recordings often starting later (mostly after 2017). For more detailed statistics, refer to the project page\footnote{\url{https://github.com/aizhiqi-work/voxaging}}.

\subsection{Collection pipeline}
\begin{figure}[t]
  \centering
  \includegraphics[width=\linewidth]{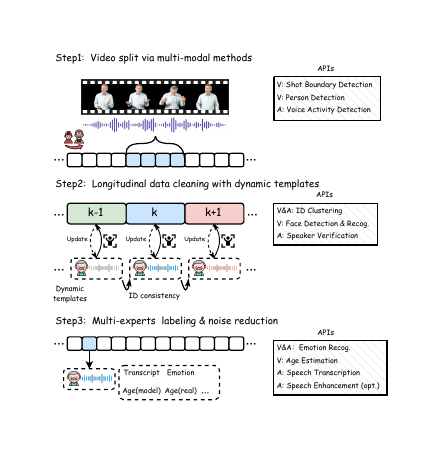}
  \caption{Illustration of the collection pipeline.}
  \label{fig:datapipeline}
\end{figure}

Our data cleaning process differs from traditional methods \cite{nagrani17_interspeech, li23y_interspeech} that rely on a single static template, as we place greater emphasis on the impact of individual aging on facial and voice features. To address this, we employ dynamic templates in the cleaning process to account for the aging of facial appearance and voice characteristics, as illustrated in Figure \ref{fig:datapipeline}. The entire cleaning process is divided into three steps:
        
\begin{itemize}
    \item \textbf{Step 1.} Video split via multi-modal methods.
    We segment videos into clips using multi-modal methods, including shot boundary detection\footnote{\url{https://www.scenedetect.com}} to identify scene transitions, YOLO-world \cite{cheng2024yolo} for person detection, and voice activity detection \cite{gao23g_interspeech} to isolate speech segments. The intersection of visual and audio boundaries is then calculated to define each segment.
    \item \textbf{Step 2.} Longitudinal data cleaning with dynamic templates.  
    We employ dynamic templates for data cleaning and use face recognition \cite{deng2018arcface} and speaker verification \cite{desplanques20_interspeech} models to extract feature representations for each segment. Then, we apply the DBSCAN clustering algorithm to group similar speaker identities from different periods, removing noisy data and ensuring ID consistency. Finally, these dynamic templates are used to refine the cleaning process for each time segment.
    
    \item \textbf{Step 3.} Multi-experts labeling \& noise reduction.
    We utilize multiple expert models to annotate and refine the cleaned data. Specifically, we employ a speech transcription model \cite{radford2023robust}, a multi-modal emotion recognition model \cite{serengil2021lightface, ma24b_interspeech}, and an age estimation model \cite{serengil2021lightface} to label the data. The age estimation model is particularly crucial, as it assigns age groups to each ID. During the initial data collection, we could only determine the timespan of each video, without knowing the user's actual age. Finally, we apply speech enhancement models \cite{liu2022separate} to the high-quality data for noise reduction, further improving the accuracy of age analysis.
\end{itemize}

\section{Experiments}
\subsection{Data setting}
As shown in Table \ref{tab:setting}, the data settings of VoxAging include "X-Independent" and "X-Dependent" configurations. 
\begin{itemize}
    \item The "X-Independent" setting consists of two subsets: VoxAging-EN (English speakers) and VoxAging-ZH (Mandarin speakers). This setup investigates the impact of aging on speaker verification systems. VoxAging-EN is divided into 11 time spans (0 to 10 years), while VoxAging-ZH is divided into 5 time spans (0 to 4 years).
    \item The "X-Dependent" setting, using VoxAging-EN, explores the effects of age group (VoxAging-AgeGroup) and gender (VoxAging-Gender) on speaker aging. The dataset is divided into 5 age groups. It also includes 114 male and 112 female speakers. Both analyses cover 6 time spans: 0, 2, 4, 6, 8, and 10 years.
\end{itemize}

\subsection{Model setting}

As shown in Table \ref{tab:eer}, to investigate the impact of speaker aging on state-of-the-art speaker recognition models, we first employed the face recognition model ArcFace \cite{deng2018arcface} as a baseline for aging. Subsequently, we evaluated seven advanced speaker recognition models \cite{zheng20233d}, which demonstrated varying performances on the VoxCeleb dataset \cite{nagrani17_interspeech}. These models include RDINO \cite{rdino}, TDNN \cite{snyder2018x}, SDPN \cite{chen2024sdpn}, ECAPA-TDNN \cite{desplanques20_interspeech}, CAM++ \cite{wang2023cam++}, ERes2Net \cite{eres2net}, and ERes2Net-large \cite{eres2net}. Among these, the best-performing model was ERes2Net-large\footnote{\url{https://github.com/modelscope/3D-Speaker}}, achieving an EER of 0.57\% on Vox-O\footnote{\url{https://www.modelscope.cn/models/iic/speech_eres2net_large_sv_en_voxceleb_16k}}.

\begin{table}[!htp]
\centering
\caption{Data setting for VoxAging.}
\resizebox{\linewidth}{!}{
\begin{tabular}{@{}llcl@{}}
\toprule
\multicolumn{2}{c}{Setting} & \# of Spks & \# of Trails \\ \midrule
\multicolumn{4}{l}{\color[HTML]{C0C0C0} {\textit{X-Independent}}} \\ \midrule
VoxAging-EN & Cross-Age & 226 & 1.1M \\
VoxAging-ZH & Cross-Age & 67 & 0.5M \\ \midrule
\multicolumn{4}{l}{\color[HTML]{C0C0C0} {\textit{X-Dependent (EN)}}} \\ \midrule
\multirow{5}{*}{VoxAging-AgeGroup} & \textless{}30 & 92 & \multirow{5}{*}{3.0M} \\
 & 30$\sim$40 & 77 &  \\
 & 40$\sim$50 & 31 &  \\
 & 50$\sim$60 & 16 &  \\
 & \textgreater{}60 & 10 & \\ \midrule
\multirow{2}{*}{VoxAging-Gender} & Male & 114 & \multirow{2}{*}{1.2M} \\ 
 & Famale & 112 &  \\ \bottomrule
\end{tabular}
}
\label{tab:setting}
\end{table}

\section{Results}
\subsection{Impact of speaker aging on advanced speaker verification systems}

Table \ref{tab:eer} shows the impact of speaker aging on advanced speaker verification systems. These models perform differently on the general test set Vox-O \cite{nagrani17_interspeech}, and we use Equal Error Rate (EER) to evaluate the effect of aging on the VoxAging-EN and VoxAging-ZH subsets. As the time span increases, the EER of the speaker verification system deteriorates, indicating that the speaker recognition accuracy declines over time. Additionally, we use the face recognition model (ArcFace \cite{deng2018arcface}) as the baseline for aging analysis. Compared to the speaker verification model, ArcFace demonstrates greater robustness to facial aging, delivering exceptional performance. However, despite this robustness, recognition accuracy still declines over time, with the EER rising from 0.31\% to 1.52\%.

\begin{table*}[!htp]
\centering
\caption{Impact of speaker aging on advanced speaker verification systems.}
\resizebox{\linewidth}{!}{
\begin{tabular}{@{}lccccccccccccclcccccc@{}}
\toprule
 &
   &
  \multicolumn{12}{c}{VoxAging-EN EER(\%)↓} &
   &
  \multicolumn{6}{c}{VoxAging-ZH EER(\%)↓} \\ \cmidrule(lr){3-14} \cmidrule(l){16-21} 
\multirow{-2}{*}{Model} &
  \multirow{-2}{*}{Vox-O} &
  0 &
  1 &
  2 &
  3 &
  4 &
  5 &
  6 &
  7 &
  8 &
  9 &
  10 &
  $\Delta$ &
   &
  0 &
  1 &
  2 &
  3 &
  4 &
  $\Delta$ \\ \midrule
\multicolumn{21}{l}{{\color[HTML]{C0C0C0} \textit{Face Modality}}} \\ \midrule
ArcFace $\cite{deng2018arcface}$ &
  - &
  0.31 &
  0.38 &
  0.55 &
  0.72 &
  0.75 &
  1.00 &
  1.23 &
  1.23 &
  1.42 &
  1.56 &
  1.52 &
  1.21 &
   &
  0.52 &
  0.58 &
  0.62 &
  0.75 &
  0.82 &
  0.30 \\ \midrule
\multicolumn{21}{l}{{\color[HTML]{C0C0C0} \textit{Speech Modality}}} \\ \midrule
RDINO $\cite{rdino}$ &
  3.16 &
  6.19 &
  6.42 &
  6.82 &
  7.25 &
  7.27 &
  7.61 &
  7.92 &
  8.09 &
  8.51 &
  8.72 &
  9.17 &
  2.98 &
   &
  17.70 &
  19.31 &
  19.68 &
  20.32 &
  20.16 &
  2.46 \\
TDNN $\cite{snyder2018x}$ &
  2.22 &
  6.43 &
  6.59 &
  6.85 &
  7.15 &
  7.15 &
  7.37 &
  7.53 &
  7.65 &
  7.81 &
  8.23 &
  8.36 &
  1.93 &
   &
  \textbf{9.06} &
  \textbf{9.82} &
  \textbf{10.63} &
  \textbf{11.58} &
  \textbf{11.78} &
  2.72 \\
SDPN $\cite{chen2024sdpn}$ &
  1.88 &
  \textbf{2.86} &
  \textbf{2.91} &
  \textbf{3.00} &
  \textbf{3.20} &
  \textbf{3.18} &
  \textbf{3.20} &
  \textbf{3.31} &
  \textbf{3.45} &
  \textbf{3.55} &
  \textbf{3.78} &
  \textbf{3.73} &
  \textbf{0.87} &
   &
  13.98 &
  15.76 &
  16.50 &
  17.15 &
  16.34 &
  $\underline{2.36}$ \\
ECAPA-TDNN $\cite{desplanques20_interspeech}$ &
  0.86 &
  4.07 &
  4.16 &
  4.47 &
  4.49 &
  4.52 &
  4.53 &
  4.66 &
  4.86 &
  5.04 &
  5.27 &
  5.36 &
  1.29 &
   &
  11.15 &
  12.77 &
  14.02 &
  14.88 &
  14.75 &
  3.60 \\
ERes2Net $\cite{eres2net}$ &
  0.83 &
  3.02 &
  3.26 &
  3.40 &
  3.61 &
  3.56 &
  3.57 &
  3.67 &
  3.75 &
  3.87 &
  4.08 &
  4.20 &
  1.18 &
   &
  $\underline{10.30}$ &
  $\underline{11.08}$ &
  $\underline{12.18}$ &
  $\underline{12.57}$ &
  $\underline{12.43}$ &
  \textbf{2.13} \\
CAM++ $\cite{wang2023cam++}$ &
  $\underline{0.65}$ &
  3.72 &
  3.94 &
  4.13 &
  4.19 &
  4.31 &
  4.19 &
  4.29 &
  4.46 &
  4.64 &
  4.76 &
  4.80 &
  1.08 &
   &
  12.53 &
  14.37 &
  15.91 &
  16.44 &
  16.37 &
  3.84 \\
ERes2Net-large $\cite{eres2net}$ &
  \textbf{0.57} &
  $\underline{2.89}$ &
  $\underline{3.05}$ &
  $\underline{3.11}$ &
  $\underline{3.24}$ &
  $\underline{3.22}$ &
  $\underline{3.24}$ &
  $\underline{3.37}$ &
  $\underline{3.47}$ &
  $\underline{3.66}$ &
  $\underline{3.80}$ &
  $\underline{3.87}$ &
  $\underline{0.98}$ &
   &
  10.52 &
  11.87 &
  12.91 &
  13.74 &
  13.72 &
  3.20 \\ \bottomrule
\end{tabular}
}
\label{tab:eer}
\end{table*}

In VoxAging-EN, RDINO \cite{rdino} and TDNN \cite{snyder2018x} show relatively poor performance, as reflected by their higher initial EERs and deterioration rates of 2.98\% and 1.93\%, respectively. In contrast, ECAPA-TDNN \cite{desplanques20_interspeech}, ERes2Net \cite{eres2net}, CAM++ \cite{wang2023cam++}, and ERes2Net-Large \cite{eres2net} exhibit lower initial EERs and slower deterioration rates, suggesting that improving the performance of speaker recognition models can enhance their robustness against speaker aging. In VoxAging-ZH, all models display generally higher initial EERs and greater deterioration rates (significantly higher than in VoxAging-EN), but the overall trend remains consistent with VoxAging-EN.

However, there are some special cases. In VoxAging-EN, the initial EER and deterioration rate of SDPN \cite{chen2024sdpn} are comparable to those of ERes2Net-Large. In VoxAging-ZH, the initial EER of TDNN is relatively low, at only 9.06\%.

\subsection{Speaker similarity scores over time}
\begin{figure}[t]
  \centering
  \includegraphics[width=\linewidth]{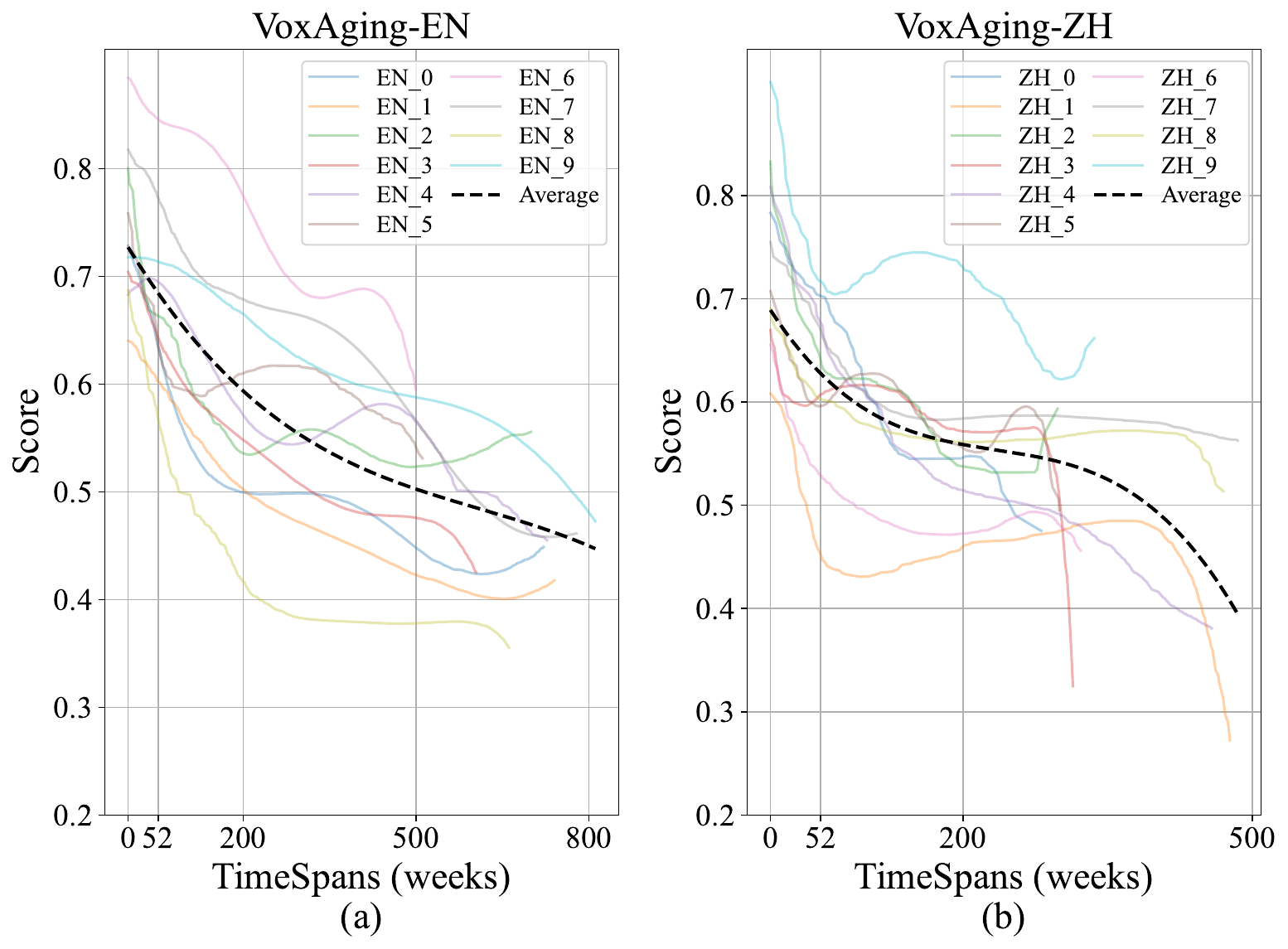}
  \caption{Speaker similarity scores over time in VoxAging. Dashed black line indicates the average aging trend.}
  \label{fig:sss}
\end{figure}

Figure \ref{fig:sss} shows the trend of speaker similarity scores over time in VoxAging, where embeddings were extracted using ECAPA-TDNN \cite{desplanques20_interspeech}. We randomly selected 10 English and 10 Mandarin speakers from the dataset and analyzed speaker similarity using a cubic polynomial fitting method over a weekly time span. The results show that speaker similarity decreases over time from the point of enrollment. This decline is caused by age-related changes in the speakers' voices, which emphasizes a key factor affecting the performance of speaker verification systems.

In Figure \ref{fig:sss}, the black dashed line represents the average trend of the speaker similarity score decline. It is clearly evident that there is a difference in the decay rate of speaker similarity between English and Mandarin. For the English average trend, it takes about 500 weeks ($\sim$10 years) for the speaker similarity to fall below the 0.5 threshold, while for the Mandarin average trend, it takes about 400 weeks ($\sim$8 years) for the speaker similarity to fall below the 0.5 threshold.

\subsection{Impact of age group and gender on speaker aging}
Table \ref{tab:agg} shows the impact of age group and gender on speaker aging, with embeddings extracted using ERes2Net-Large \cite{eres2net}. In all age groups, the performance of the speaker verification system deteriorates with age. In VoxAging-AgeGroup, the initial EER for the young age group ($\textless{30}$ years) is relatively high, reaching 5.24\% at the 10-year mark. The initial EER for the 30$\sim$40 and 40$\sim$50 age groups is lower than that of the young group, but the aging effect is more pronounced, with deterioration rates of 1.50\% and 1.67\%, respectively. The initial EER for the 50$\sim$60 age group is similar to that of the 40$\sim$50 group, but the deterioration is slower (1.17\%). For those over 60 years old, the aging effect is the least pronounced, and the overall EER remains relatively stable, with a deterioration rate of 0.30\%. Overall, the experiment shows that age-related voice changes are particularly significant in the 40$\sim$50 age group.


\begin{table}[!htp]
\centering
\caption{The impact of age group and gender on speaker aging.}
\resizebox{\linewidth}{!}{
\begin{tabular}{@{}ccccccccc@{}}
\toprule
\multicolumn{2}{c}{\multirow{2}{*}{Setting}}          & \multicolumn{7}{c}{EER(\%)↓}                                         \\ \cmidrule(l){3-9} 
\multicolumn{2}{c}{}                                  & 0    & 2    & 4    & 6    & 8    & 10   & $\Delta$ \\ \midrule
\multirow{5}{*}{AgeGroup}          & $\textless{30}$  & 4.12 & 4.58 & 4.64 & 4.76 & 5.14 & 5.24 & 1.12                       \\
                                   & 30$\sim$40       & 3.43 & 3.63 & 3.84 & 4.20 & 4.58 & 4.93 & 1.50                       \\
                                   & 40$\sim$50       & 2.57 & 2.70 & 3.10 & 3.26 & 3.62 & 4.24 & \textbf{1.67}                       \\
                                   & 50$\sim$60       & 2.58 & 2.70 & 2.90 & 2.96 & 3.33 & 3.75 & 1.17                       \\
                                   & $\textgreater{60}$ & 2.82 & 2.91 & 2.74 & 2.91 & 3.22 & 3.12 & 0.30                       \\ \midrule
\multirow{2}{*}{Gender}   & Male             & 3.54 & 3.71 & 3.77 & 3.75 & 3.92 & 4.09 & 0.55                       \\
                                   & Female           & 4.15 & 4.77 & 5.05 & 5.50 & 6.02 & 6.77 & \textbf{2.62}                       \\ \bottomrule
\end{tabular}
}
\label{tab:agg}
\end{table}

In VoxAging-Gender, it is clear that male speakers have lower initial EER values than female speakers. Additionally, both genders exhibit similar trends, with EER values increasing over time. The deterioration is more pronounced in the female group, with a deterioration rate of 2.62\%, reaching an EER of 6.77\% at the 10-year mark, higher than the male group (4.09\%). This suggests that age-related voice changes may have a more noticeable impact on female group in VoxAging.

\section{Conclusions}

In this paper, we present VoxAging, a large-scale longitudinal 
dataset. It includes recordings from 293 speakers (226 English and 67 Mandarin) over a span of 17 years, totaling 7,522 hours, with weekly samples. Our analysis of speaker aging reveals that the performance of speaker verification systems deteriorates with age. Improving the performance of speaker recognition models can enhance their resistance to speaker aging. Additionally, speaker similarity scores significantly declines over time. The impact of age and gender on speaker aging shows that 40$\sim$50 age group and female group exhibit more pronounced voice deterioration.

\bibliographystyle{IEEEtran}
\bibliography{main}

\end{document}